\documentclass[conference]{IEEEtran}
\usepackage{graphicx}
\usepackage{epstopdf}
\usepackage{amsthm,amsmath,amssymb}
\usepackage{mathrsfs}
\usepackage{setspace}
\usepackage{autobreak}
\usepackage{amsmath}
\usepackage{booktabs}
\usepackage{enumerate}
\usepackage{cite}

\usepackage[
top    = 1.81cm,
bottom = 1.05in,
left   = 0.6 in,
right  = 0.6 in]{geometry}

\newtheorem{proposition}{Proposition}

\ifCLASSINFOpdf
\else
\fi
\begin{document}
%
\title{Resource Allocation in IRSs Aided MISO-NOMA Networks: A Machine Learning Approach}
%
%
%

\author{\IEEEauthorblockN{ Xinyu~Gao\IEEEauthorrefmark{1}, Yuanwei~Liu\IEEEauthorrefmark{1}, Xiao~Liu\IEEEauthorrefmark{1}, and Zhijin~Qin\IEEEauthorrefmark{1}} 
\IEEEauthorblockA{\IEEEauthorrefmark{1} Queen Mary University of London, London, UK\\
 }}

\maketitle

\begin{abstract}
A novel framework of intelligent reflecting surface (IRS)-aided multiple-input single-output (MISO) non-orthogonal multiple access (NOMA) network is proposed, where a base station (BS) serves multiple clusters with unfixed number of users in each cluster. The goal is to maximize the sum rate of all users by jointly optimizing the passive beamforming vector at the IRS, decoding order and power allocation coefficient vector, subject to the rate requirements of users. In order to tackle the formulated problem, a three-step approach is proposed. More particularly, a long short-term memory (LSTM) based algorithm is first adopted for predicting the mobility of users. Secondly, a K-means based Gaussian mixture model (K-GMM) algorithm is proposed for user clustering. Thirdly, a deep Q-network (DQN) based algorithm is invoked for jointly determining the phase shift matrix and power allocation policy. Simulation results are provided for demonstrating that the proposed algorithm outperforms the benchmarks, while the performance of IRS-NOMA system is better than IRS-OMA system.
\end{abstract}

\vspace{-0.2cm}

%
\IEEEpeerreviewmaketitle

\section{Introduction}
%
%
%
%
\IEEEPARstart{W}{ith} the increasing demand for large capacity in wireless networks, the conventional multiple access schemes cannot guarantee the quality of connectivity. Therefore, pursuing spectrum efficiency has become the leading focus point in wireless networks, especially in the fifth-generation (5G) era where data volume and access volume are exploding. Although various techniques have been invoked for improving spectrum efficiency, such as large-scale multiple-input multiple-output (MIMO)\cite{IEEEhowto:Walton}, millimeter-wave communications\cite{IEEEhowto:Rappaport}, ultra-massive and ubiquitous wireless connectivity, which are the goals of next-generation wireless networks, are still far from realized. Intelligent reflecting surfaces (IRSs), which have the capability of proactively modifying the wireless communication channels by controlling a large number of passive reflective elements, are recognized as a promising technique to enhance both spectrum efficiency and energy efficiency of wireless networks. Additionally, in order to further improve spectrum efficiency and user connectivity of IRS-aided wireless networks, power domain non-orthogonal multiple access (NOMA) technique\cite{IEEEhowto:Liu} can be leveraged, whose core idea is to superimpose the signals of two users at different powers for exploiting the spectrum more efficiently by opportunistically exploring the users' different channel conditions.

\par
The aid of IRSs, both spectrum efficiency and energy efficiency of wireless networks have witnessed significant improvement. The authors in \cite{IEEEhowto:Wu} jointly designed and implemented a novel IRS-aided hybrid wireless network, which showed that the IRS can be useful to achieve significant performance enhancement in typical wireless networks, compared to the traditional networks comprising active components only. In \cite{IEEEhowto:Yu}, a jointly active beamforming and passive beamforming design algorithm was proposed for physical layer security in wireless networks. By invoking NOMA technique in IRS-aided wireless networks, spectrum efficiency can be further enhanced when comparing to the conventional OMA schemes, such as Time-division multiple access (TDMA)\cite{IEEEhowto:Jindal} and Frequency-division multiple access (FDMA)\cite{IEEEhowto:Myung}. Mu \emph{et al.}\cite{IEEEhowto:Mu} exploited IRS-aided NOMA system and developed a novel algorithm by utilizing the sequential rank-one constraint relaxation approach to find a locally optimal rank-one solution. Fu \emph{et al.} \cite{IEEEhowto:Fu} considered jointly optimizing the transmit beamformers at the base station (BS) and the phase shift matrix at the IRS for a IRS-empowered NOMA network, which proved that performance gain was achieved. Reinforcement learning (RL) has shown great potentials to revolutionize communication systems. Additionally, RL was proved to be capable of tackling dynamic environment in IRS-aided wireless networks. In order to obtain the beneficial of NOMA technique, a novel framework for the deployment and passive beamforming design of a IRS with the aid of NOMA technology was proposed in \cite{IEEEhowto:Liu2}. Cui \emph{et al.}\cite{IEEEhowto:Cui} developed a K-means-based online user clustering algorithm to reduce the computational complexity and derive the optimal power allocation policy in a closed form by exploiting the successive decoding feature.

\par
Although the aforementioned research contributions have laid a foundation on solving challenges in IRS-aided wireless networks and on leveraging NOMA for improving the spectrum-efficiency of networks, the dynamic environment derived from the movement of ground mobile users is ignored in the previous research contributions. Before fully reap the advantages of IRSs and NOMA technique, how to design the phase shift matrix of the IRS and resource allocation policy based on the mobility information of users is still challenging. In contrast to the conventional MIMO-NOMA system, additional decoding rate conditions need to be satisfied to guarantee successful successive interference cancellation (SIC) in IRS-NOMA systems. Additionally, both the active beamforming and passive phase shift design affect the decoding order among users and user clustering, which makes the decoding order design, user clustering and passive beamforming design highly coupled.
\par
Sparked by the above background, we aim to find the maximum sum rate in the downlink IRS-aided MISO-NOMA network. Our contributions are summarized as follows: 1) We propose a novel framework for IRS-NOMA aided wireless network, where an IRS is employed to enhance spectrum efficiency by proactively reflecting the incident signals. 2) We conceive a long short-term memory (LSTM) algorithm based on the concept of rejection method for randomly generating users' intial positions in a fixed range and predicting future positions of users. 3) We adopt a K-means based Gaussian mixture model (K-GMM) algorithm for dynamic user clustering. 4) We demonstrate that the proposed DQN algorithm is capable of solving the joint phase shift design and resource allocation problem in IRS-aided wireless networks. Additionally, IRS-NOMA scheme outperforms IRS-OMA scheme in terms of sum rate.




\vspace{-0.25cm}
\section{System Model and Problem Formulation}
\vspace{-0.1cm}
\subsection{System Model}
\par
\begin{figure}[ht]
  \centering
  \includegraphics[height=1.6in,width=3.4in]{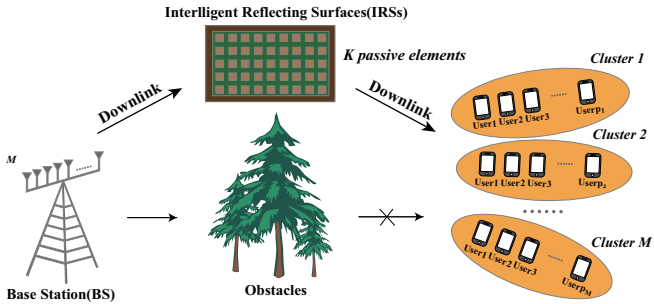}
  \caption{Illustration of the IRS-aided MISO-NOMA network.}
  \label{system_model}
  \end{figure}

As shown in Fig.~\ref{system_model}, we consider a downlink multi-cluster system, where the BS is equipped with \emph{M} transmitting antennas and a fixed number \emph{$\Lambda$} of single antenna users are served by the US, all users are partitioned into \emph{M} clusters where the number of users in each cluster is \emph{$p_{m}$}. Furthermore, in order to improve the spectrum efficiency, in each cluster, NOMA technology is applied for transmission. It is worth noting that the direct transmit link between the BS and users is blocked by obstacles. To enhance the quality-of-service (QoS), an IRS with \emph{K} low-cost passive elements is employment to assist the MISO-NOMA network.
\setlength{\abovecaptionskip}{-0.2cm}
\par
In the light of the system description, the baseband equivalent channels from the BS to IRS and the IRS to user \emph{p} in the cluster \emph{m} are denoted by \pmb{\emph{G}} $\in$ $\mathbb {C}$$^{K \times M}$, \emph{$\pmb{h}_{m,p}^{H}$} $\in$ $\mathbb {C}$$^{1 \times K}$, respectively, while \emph{u}$_{m,p}$ denotes the \emph{p}-th user in \emph{m}-th cluster. Then, assume \emph{$\phi_{k} = \beta_{k}e^{j\theta_{k}}$} denotes the reflection coefficient of \emph{k}-th element, while \emph{$\beta_{k}$} and \emph{$\theta_{k}$} denote the amplitude and phase of \emph{k}th element in the IRS. Besides, denote \emph{$\pmb{\Theta} = [\phi_{1},\phi_{2},\cdots,\phi_{k}]$} as the reflection coefficients matrix of the IRS, while \emph{$\phi_{k}$}, \emph{k=1,2,3,$\cdots$,K} as the diagonal element. Therefore, the received signal at \emph{$u_{m,p}$} is expressed as
\vspace{-0.18cm}
\begin{align}\label{1}
y_{m,p} = \pmb{h}_{m,p}^{H} \pmb{\Theta} \pmb{G} \sum\limits_{m=1}^{M}\pmb{\omega}_{m} x_{m} + n_{m,p},
\end{align}
where \emph{$x_{m}$} denotes the received signal of cluster \emph{m} transmitted from the BS, \emph{n$_{m,p}$} $\sim$ $\mathcal{CN}$(0,\emph{$\sigma$$^{2}$}) indicates the additive white Gaussian noise (AWGN) at user \emph{p} of cluster \emph{m} with zero mean and variance \emph{$\sigma$$^{2}$}, and using \emph{$\pmb{\omega}_{m}$} represents the corresponding beamforming vector for the \emph{m}-th cluster. We use \emph{$s_{m,p}$} to denote the required signal from \emph{$u_{m,p}$}, then the mixed signal received by the \emph{k}-th cluster is formulated as
\vspace{-0.18cm}
\begin{align}\label{2}
x_{m} = \sum\limits_{p=1}^{p_{m}} \alpha_{m,p}s_{m,p},
\end{align}
where the \emph{$\alpha_{m,p}$} and \emph{$s_{m,p}$} as the power allocation coefficient and the transmitted information for user \emph{$u_{m,p}$}, respectively, while satisfied \emph{$\sum_{p=1}^{P_{m}}\alpha_{m,p} = 1$}.
Then we use the \emph{B} as the resolution bits, so the discrete one is expressed as
\vspace{-0.18cm}
\begin{align}\label{3}
\mid \beta_{k} \mid^{2} = 1, \theta_{k}\in \{ \frac{2\pi n}{2^{B}},n=0,1,2,\cdots,2^{B}-1\}.
\end{align}
\par
Based on \eqref{1}, the received signal of user \emph{$u_{m,p}$} is given by
\vspace{-0.18cm}
\begin{align}\label{4}
y_{m,p} = &\underbrace{\pmb{h}_{m,p}^{H} \pmb{\Theta} \pmb{G} \pmb{\omega}_{m} \alpha_{m,p}s_{m,p}}_{{\rm Desired\ signal}} + \underbrace{\pmb{h}_{m,p}^{H} \pmb{\Theta} \pmb{G} \pmb{\omega}_{m} \sum\limits_{\lambda=1,\lambda \neq p}^{p_{m}}\alpha_{m,\lambda}s_{m,\lambda}}_{{\rm intra-cluster\ interference}} \nonumber\\
\vspace{-0.1cm}
&+ \underbrace{\pmb{h}_{m,p}^{H} \pmb{\Theta} \pmb{G} \sum\limits_{\gamma=1, \gamma\neq m}^{M}\pmb{\omega}_{\gamma} s_{\gamma}}_{{\rm inter-cluster\ interference}} + n_{m,p},
\end{align}
where \emph{$\pmb{h}_{m,p}^{H} \pmb{\Theta} \pmb{G} \pmb{\omega}_{m}\sum_{\lambda=1,\lambda \neq p}^{p_{m}} \alpha_{m,p}s_{m,p}$} denotes the intra-cluster interference from other users in the same cluster and \emph{$\pmb{h}_{m,p}^{H} \pmb{\Theta} \pmb{G}\sum_{\gamma=1, \gamma\neq m}^{M}\pmb{\omega}_{\gamma} s_{\gamma}$} represents the inter-cluster interference from other clusters. For beamforming matrix \emph{$\omega_{m}$}, the zero-forcing (ZF)-based linear method is applied, the corresponding ZF pre-coding constraints are expressed as
\vspace{-0.18cm}
\begin{align}\label{5}
  \left\{
    \begin{array}{lr}
      \pmb{h}_{\gamma}^{H}\pmb{\Theta}\pmb{G}\pmb{\omega}_{m} = 0, &  \\
      \pmb{h}_{m}^{H}\pmb{\Theta}\pmb{G}\pmb{\omega}_{m} = 1,
    \end{array}
  \right.
\end{align}
where \emph{$\pmb{h}_{\gamma}$} represents the combined channel of the \emph{$\gamma$}-th cluster. Then, we denote \emph{$\pmb{h}^{H} = \pmb{h}_{M}^{H}\Theta G$}, in which \emph{$\pmb{h}_{M}^{H} = [h_{1,p},h_{2,p},\cdots,h_{M,p}]^{H}$}. Therefore, the optimal transmit pre-coding beamforming vector is given by
\vspace{-0.18cm}
\begin{align}\label{6}
\pmb{W} = [\pmb{\omega}_{1},\pmb{\omega}_{2},\cdots,\pmb{\omega}_{M}] = \pmb{h}(\pmb{h}^{H}\pmb{h})^{-1}.
\end{align}
\par
Then, we denote \emph{$\pmb{h}_{m,p}^{H} \pmb{\Theta} \pmb{G} = \pmb{\upsilon}^{H}\pmb{\Phi_{m,p}}$}, where \pmb{$\Phi$} = diag($\pmb{h}_{m,p}^{H}$)\pmb{G}, $\pmb{\upsilon} = [\upsilon_{1}, \upsilon_{2}, \cdots, \upsilon_{k}]^{H}$ where $\upsilon_{k} = \emph{e}^{j\theta_{k}}$, and \emph{$\pmb{\Gamma}_{\gamma}=\sum_{\gamma=1, \gamma\neq m}^{M}\pmb{\omega}_{\gamma}$}, so the signal-to-interference-plus-noise ratio (SINR) of \emph{$u_{m,p}$} is given by
\vspace{-0.18cm}
\begin{align}\label{7}
\tau_{m,p} = \frac{\mid\pmb{\upsilon}^{H}\pmb{\Phi}_{m,p}\pmb{\omega}_{m}\alpha_{m,p}\mid^2}{\sum\limits_{\lambda=1,\lambda \neq p}^{p_{m}}\mid\pmb{\upsilon}^{H}\pmb{\Phi}_{m,p}\pmb{\omega}_{m}\alpha_{m,\lambda}\mid^2+\mid\pmb{\upsilon}^{H}\pmb{\Phi}_{m,p}\pmb{\Gamma}_{\gamma}\mid^2+\delta_{m}^{2}}.
\end{align}
\par
Similarly, the SINR for \emph{$u_{m,q}$} to decode the received signal \emph{$s_{m,p}$} is expressed as
\vspace{-0.18cm}
\begin{align}\label{8}
\tau_{m,q,p} = \frac{\mid\pmb{\upsilon}^{H}\pmb{\Phi}_{m,q}\pmb{\omega}_{m}\alpha_{m,p}\mid^2}{\sum\limits_{\lambda=1,\lambda \neq p}^{p_{m}}\mid\pmb{\upsilon}^{H}\pmb{\Phi}_{m,q}\pmb{\omega}_{m}\alpha_{m,\lambda}\mid^2+\mid\pmb{\upsilon}^{H}\pmb{\Phi}_{m,q}\pmb{\Gamma}_{\gamma}\mid^2+\delta_{m}^{2}}.
\end{align}
\par
It is worth pointing out that all the users have to meet QoS requirement and to guarantee success SIC under a given decoding order. Therefore, the following constraint has to be satisfied
\vspace{-0.18cm}
\begin{align}\label{9}
\tau_{m,p}\geq\tau_{m,\tilde{p}},
\end{align}
where \emph{$\tau_{m,\tilde{p}}$} represents the minimum received SINR that the weakest user \emph{$u_{m,\tilde{p}}$} has to achieve. Therefore, among all the users, the success SIC decoding is subjected to
\vspace{-0.18cm}
\begin{align}\label{10}
R_{m,a \rightarrow m,b} \geq R_{m,b \rightarrow m,b},
\end{align}
when decording order \emph{$\Omega_{m,a}>\Omega_{m,b}$}, for users \emph{a} and \emph{b} in any cluster \emph{m}.
\begin{proposition}\label{proposition 1}
For any two users a and b in cluster m, given the optimal decoding order, $R_{m,a \rightarrow m,b} \geq R_{m,b \rightarrow m,b}$ is the necessary constraint for $R_{m,a \rightarrow m,b} \geq R_{m,b \rightarrow m,\tilde{b}}$.
\begin{proof}
  See Appendix~A.
\end{proof}
\end{proposition}
\textbf{Proposition 1} indicates that when the optimal decoding order of NOMA is given, the constraint \eqref{9} can be removed, while the performance of NOMA networks will not be affected.

\vspace{-0.1cm}
\subsection{Problem Formulation}
\vspace{-0.05cm}
In this paper, we will design a novel protocol for achieving maximum sum rate of all users in the IRS-aided MISO-NOMA network by jointly optimizing the passive beamforming vector \emph{$\pmb{\upsilon}$} at the IRS, decoding order \emph{$\pmb{\Omega}$} and power allocation coefficient vector \emph{$\pmb{\alpha}$}, subject to the rate requirements at \emph{$\Lambda$} users. Thus, the optimization problem is formulated as
\vspace{-0.18cm}
\begin{align}
  \max_{\pmb{\Omega},\pmb{\upsilon},\pmb{\omega},\pmb{\alpha}} \hspace*{1em}&\sum\limits_{m=1}^{M}\sum\limits_{p=1}^{P_{m}}R_{m,p} \label{11}\\
  {\rm s.t.} \hspace*{1em}&  R_{m,a \rightarrow m,b} \geq R_{m,b \rightarrow m,b} \tag{\ref{11}{a}}, \label{11a}\\
  &\sum\limits_{m=1}^{M} \mid\mid \pmb{\omega} \mid\mid^{2} \leq \mathcal{P}, \tag{\ref{11}{b}} \label{11b}\\
  &\mid \beta_{k} \mid^{2} = 1, \theta_{k}\in \{ \frac{2\pi n}{2^{B}},n=0,1,\cdots,2^{B}-1\} \tag{\ref{11}{c}}, \label{11c}\\
  &\pmb{\Omega} \in \Pi \tag{\ref{11}{d}}, \label{11d}
\end{align}
where \emph{$\pmb{\omega} = \{\pmb{\omega}_{1},\pmb{\omega}_{2},\cdots,\pmb{\omega}_{M}\}$}, \emph{$\pmb{\alpha} = \{\pmb{\alpha}_{1},\pmb{\alpha}_{2},\cdots,\pmb{\alpha}_{m}\}$}, \emph{$\pmb{\alpha}_{m} = \{\alpha_{m,1},\alpha_{m,2},\cdots,\alpha_{m,p}\}$}, \emph{$\mathcal{P}$} represents the transmit power. Constraint (\ref{11a}) guarantees that the SIC can be be performed successfully. Constraint (\ref{11b}) is the total transmission power constraint. Constraint (\ref{11c}) represents the considered IRS assumption. Finally, constraint (\ref{11d}) denotes the set of all the possible decoding orders. Since users are considered as roaming continuously, the optimal decoding order has to be re-determined at each timeslot for success SIC, which is naturally a dynamic problem. Since the conventional convex optimization is non-trivial to tackle the formulated dynamic problem, machine learning algorithms will be introduced in the following sections.

\vspace{-0.2cm}
\section{Proposed solutions}
\vspace{-0.05cm}
\subsection{Positions predicting based on deep learning method}
\subsubsection{Training samples selection}
In our proposed model, we use the rejection method for generating training samples. Firstly, a fixed number of \emph{$N_{0}$} positions for each user is randomly generated during the intimal time period. Afterward, all these position points are served as training samples for the first LSTM training. Additionally, in order to achieve the expected results, the training samples vary over time. After each training, the LSTM model will predict \emph{N} positions.

\subsubsection{Training and prediction}
\textbf{Algorithm 1} demonstrates the use of rejection method and LSTM model to generate initial users' positions and predict their future positions.
\begin{table}[htbp]
  \vspace{-0.1cm}
    \begin{tabular}{l}
      \toprule
      \textbf{Algorithm 1} Positions generation and predictions\\
      \midrule
      \textbf{Input:} LSTM network structure\\
      \hspace*{0.5em} \textbf{Initialization:} Parameters of LSTM network\\
      \hspace*{0.5em} Maximum training samples \emph{$N_{max}$}, all \emph{$\Lambda$} users\\
      \hspace*{0.5em} Generate \emph{$N=N_{0}$} positions using rejection method\\
      \hspace*{0.5em} \textbf{Repeat:}\\
      \hspace*{1.5em} Input training sample set \emph{N}\\
      \hspace*{1.5em} Predict \emph{\pmb{pos} = N} positions\\
      \hspace*{1.5em} Set \emph{$N = N_{0} + N$} as training samples\\
      \hspace*{1.5em} \textbf{End}\\
      \hspace*{0.5em} \textbf{Until} \emph{$N = N_{max}$}\\
      \textbf{Return:} Matrix \emph{\pmb{pos}}\\
      \bottomrule
    \end{tabular}
    \vspace{-0.1cm}
\end{table}

\vspace{-0.05cm}
\subsection{Clustering based on K-GMM model}
\vspace{-0.05cm}
In this section, we propose a three-step user clustering approach based on K-GMM algorithm.
\subsubsection{Step 1}
All users provide their own channel state information (CSI) feedback to the transmitter and the transmitter forms a CSI set \emph{$\pmb{S}$} for all users, that is
\vspace{-0.18cm}
\begin{align}\label{12}
\pmb{S} = \{h_{1},h_{2},\cdots,h_{\Lambda}\}.
\end{align}
\par
We initialize all users into \emph{M} clusters and use \emph{$\pmb{S}_{m}$} to denote the CSI set for cluster \emph{m}. Besides, in order to normalize user channels, the channel vector of all users is defined as
\vspace{-0.18cm}
\begin{align}\label{13}
\tilde{h}_{\Lambda} = \frac{h_{\Lambda}}{\mid\mid h_{\Lambda} \mid\mid}.
\end{align}
\subsubsection{Step 2}
We randomly selects \emph{m} users as the centers of clusters, the initial value of GMM and the definition of gain difference and correlation is expressed as
\vspace{-0.18cm}
\begin{align}\label{14}
d_{a,b} = \mid\mid\tilde{h}_{a}\mid-\mid\tilde{h}_{b}\mid\mid<\rho_{1},Cor(a,b) = \dfrac{\mid\tilde{h}_{a}\cdot\tilde{h}_{b}\mid}{\mid\tilde{h}_{a}\mid\cdot\mid\tilde{h}_{b}\mid}>\rho_{2},
\end{align}
where \emph{$\rho_{1}$} and \emph{$\rho_{2}$} denote the pre-defined correlation real number thresholds while satisfying \emph{$\rho_{1},\rho_{2} \geq 0$}. Afterward, the users are roughly partitioned into \emph{M} clusters. Denote the \emph{$C_{m}$}, \emph{$\tilde{h}_{p_{m}}$} and \emph{$\pmb{S}_{m}$} as the center of \emph{m}-th cluster, a user in the \emph{m}the cluster and the initially formed user set, respectively. Therefore, the center can be reformulated
\vspace{-0.18cm}
\begin{align}\label{15}
\tilde{C}_{m} = \frac{1}{\mid C_{m}\mid} \sum\limits_{h_{p_{m}} \in \pmb{S}_{m}} \tilde{h}_{p_{m}}.
\end{align}
\subsubsection{Step 3}
The GMM model consists of \emph{m} Gaussian distributions, each Gaussian distribution is called a "component" and these components are linearly added together for any user \emph{$\lambda$}, which can be expressed as
\vspace{-0.18cm}
\begin{align}\label{16}
P(\tilde{h}_{\lambda}|\pmb{\kappa}) = \sum\limits_{m=1}^{p_{m}} \Psi_{m} p(\tilde{h}_{\lambda}|\kappa_{m}),
\end{align}
where \emph{$\Psi_{m} \geq 0, \sum\Psi_{m} = 1$} denotes the weights of each Gaussian distribution, \emph{$p(\tilde{h}_{\lambda}|\kappa_{m})$} is the probability density function of the \emph{m}-th Gaussian distribution while \emph{$\kappa_{m} = (\tilde{C}_{m}, \Psi_{m}^{2})$}, so the expression of probability density is
\vspace{-0.18cm}
\begin{align}\label{17}
p(\tilde{h}_{\lambda}|\kappa_{m}) = \frac{1}{\sqrt{2\pi}\delta_{m}}exp(-\frac{(\tilde{h}_{\lambda} - C_{m})^2}{2\delta_{m}^{2}}).
\end{align}
\par
The learning process of the GMM model is to estimate all the probability density function \emph{$p(\tilde{h}_{\lambda}|\kappa_{m})$} of \emph{M} Gaussian distributions. The probability of occurrence of each observation sample is expressed as a weighted probability of \emph{M} Gaussian distributions.
\vspace{-0.1cm}
\begin{proposition}\label{proposition 2}
  Input observation data \pmb{S} and M GMM models, iteratively converge to a small number $\epsilon$ and output parameter \pmb{$\kappa$} of all the GMM models. The parameters of GMM are derived by the EM algorithm.
 \vspace{-0.1cm}
  \begin{proof}
    See Appendix~B.
  \end{proof}
\end{proposition}
\vspace{-0.2cm}
From equation \eqref{14} to \eqref{17}, the parameters to be estimated are \emph{$\pmb{\kappa} = \{\Psi_{1}, \Psi_{2}, \cdots, \Psi_{M}; \kappa_{1}, \kappa_{2}, \cdots, \kappa_{M}\}$} and \emph{$\kappa_{M} = (\mu_{M}, \delta_{M}^{2})$}. Therefore, the \emph{3M} parameters have to be estimated in this model. The maximum likelihood estimation (MLE) method is adopted to estimate \emph{$\pmb{\kappa}$}, so that the log-likelihood function \emph{$L(\pmb{\kappa}) = logP(\tilde{h}_{\lambda}|\pmb{\kappa})$} of the observation data \emph{\pmb{S}} is maximized, which can be expressed as
\vspace{-0.18cm}
\begin{align}\label{18}
L(\pmb{\kappa}) = logP(\pmb{S}|\pmb{\kappa}) = \sum\limits_{\lambda=1}^{\Lambda}[log(\sum\limits_{m=1}^{M}\Psi_{m} p(\tilde{h}_{\lambda}|\kappa_{m})].
\end{align}
\par
The initial values of \emph{$\mu$} and \emph{$\delta^{2}$} for each Gaussian distribution are given. As mentioned above, K-means algorithm is applied to obtain the cluster center as the initial \emph{$\mu$} value. For \emph{$\delta^{2}$}, our purpose is to get the maximum value of log-likelihood function \emph{$L(\pmb{\kappa})$}, so \emph{$L(\pmb{\kappa})$} can be differentiated in a single sample as
\vspace{-0.18cm}
\begin{align}\label{19}
  d(tr(\tilde{L}(\kappa)) = -\frac{tr (\delta^{-2}d\delta^{2} - \delta^{-2})(\tilde{h} - C)(\tilde{h} - C)^{T}\delta^{-2}d\delta^{2}}{2},
\end{align}
\vspace{-0.5cm}
\begin{align}\label{20}
\frac{d(tr(\tilde{L}(\kappa))}{d\delta^{2}} = -\frac{tr(\delta^{-2} - \delta^{-2})(\tilde{h} - C)(\tilde{h} - C)^{T}\delta^{-2}}{2}.
\end{align}
\par
Thus the initialize values are given by
\vspace{-0.18cm}
\begin{align}\label{21}
\mu_{m} = \tilde{C_{m}},\delta_{m}^{2} = \frac{1}{p_{m}}\sum\limits_{\lambda=1}^{p_{m}}(\tilde{h}_{\lambda} - C_{m})(\tilde{h}_{\lambda} - C_{m})^{T}.
\end{align}
\par
The probability of user \emph{$\tilde{h}_{\lambda}$} in the \emph{m}-th Gaussian distribution is
\vspace{-0.18cm}
\begin{align}\label{22}
\chi_{\tilde{h}_{\lambda,m}} = \frac{\Psi_{m} p(\tilde{h}_{\lambda}|\kappa_{m})}{\sum\limits_{m=1}^{M} \Psi_{m} p(\tilde{h}_{\lambda}|\kappa_{m})}, \forall \lambda \in \{1,2, \cdots, \Lambda\}.
\end{align}
\par
Recalculate the parameters, we can achieve
\vspace{-0.18cm}
\begin{align}\label{23}
\tilde{\mu}_{m} = \frac{\sum\limits_{\lambda=1}^{\Lambda}\chi_{\tilde{h}_{\lambda,m}}\tilde{h}_{\lambda}}{\sum\limits_{\lambda=1}^{\Lambda}\chi_{\tilde{h}_{\lambda,m}}},\tilde{\Psi}_{m} = \frac{\sum\limits_{\lambda=1}^{\Lambda}\chi_{\tilde{h}_{\lambda,m}}}{\Lambda},
\end{align}
\vspace{-0.18cm}
\begin{align}\label{24}
\tilde{\delta}_{m}^{2} = \frac{\sum\limits_{\lambda=1}^{\Lambda}\chi_{\tilde{h}_{\lambda,m}}(\tilde{h}_{\lambda} - \mu_{m})^{2}}{\sum\limits_{\lambda=1}^{\Lambda}\chi_{\tilde{h}_{\lambda,m}}}.
\end{align}
\par
Repeat the calculation of E-step and M-step, when \emph{$\mid\mid \kappa_{t+1} - \kappa_{t} \mid\mid < \epsilon$} is satisfied, the judgment is converged, which means that the parameters of the GMM is obtained and user clustering is finished.

\vspace{-0.2cm}
\subsection{Phase shift design based on deep Q-network model}
In this section, the deep Q-network based algorithm is proposed for phase shift design of the IRS. In the DQN model, we select the BS as an agent, the BS can control resource allocation from BS to users and phase adjustment of the IRS. The BS observes the state of user clustering at each time slot. Given a state space \emph{S}, the phase shift and power allocation policies are obtained in this space. When the BS carries out an action \emph{$A_{t_{\iota}}$}, it will directly adjust the IRS to the most suitable phase at the current timeslot, which belongs to the decision-making process, representing by \emph{$\mathcal{D}$}. Accordingly, Q-function is given by
\vspace{-0.18cm}
\begin{align}\label{25}
  Q_{S_{t_{\varphi+1}},A_{t_{\iota+1}}} &= Q_{S_{t_{\varphi}},A_{t_{\iota}}} + \psi(R_{S_{t_{\varphi}},A_{t_{\iota}}} \nonumber\\
  &+\beta\cdot max \tilde{Q}_{\tilde{S}_{t_{\varphi}},\tilde{A}_{t_{\iota}}} - Q_{S_{t_{\varphi}},A_{t_{\iota}}}),
\end{align}
where \emph{$\psi$} and \emph{$\beta$} represent learning efficiency and discount parameter, respectively. Q-learning model may suffer from memory problems. In order to solve this problem, we use the Function Approximation(FA) method to introduce a function with weights \emph{$\pmb{\vartheta}$} to approximate the Q-table. Thus, the new Q value is re-calculated by
\vspace{-0.18cm}
\begin{align}\label{26}
Q^{*}(S,A) = E[R+\beta\max Q^{*}(S^{'},A^{'})|S,A],
\end{align}
where \emph{$Q^{*}(S,A)$} is the Q value function of DQN, then, the loss function is expressed as
\vspace{-0.18cm}
\begin{align}\label{27}
Loss(\pmb{\vartheta}) = \sum\limits (y - Q_{S_{t_{\varphi+1}},A_{t_{\iota+1}}, \vartheta_{t_{\epsilon+1}}}),
\end{align}
where \emph{y} represents the output value calculated by the current Q-value at the next timeslot, which is given by
\vspace{-0.18cm}
\begin{align}\label{28}
y = R_{S_{t_{\varphi}},A_{t_{\iota}}} + \beta\times max\tilde{Q}_{\tilde{S}_{t_{\varphi}},\tilde{A}_{t_{\iota}}}.
\end{align}

\vspace{-0.2cm}
\section{Simulation results}
\subsection{Positions prediction and user clustering}
\begin{figure}[t!]
  \vspace{-0.1cm}
  \begin{center}
  \includegraphics[height=1.9in,width=2.7in]{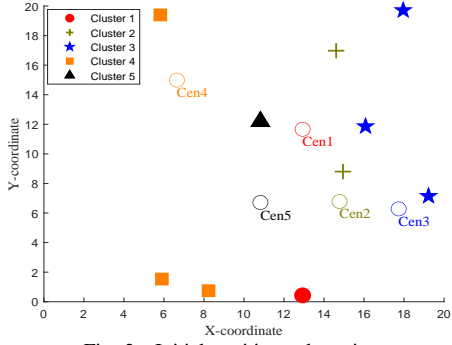}
  \caption{Initial positions clustering.}
  \label{Initial positions clustering}
  \end{center}
  \end{figure}

As discussed in the system model, we consider a total of 10 users partitioned into 5 clusters. In the light of the schemes, we randomly select 5 users as a reference point and the initial cluster center can be calculated by equation \eqref{15}. The process of iteration follows the convergence condition \emph{$\mid\mid \kappa_{t+1} - \kappa_{t} \mid\mid < \epsilon$}. It is worth noting that we set the value of \emph{$\epsilon$} as 1e-15 and IRS is deployed at the origin of coordinates and size is ignored. Fig.~\ref{Initial positions clustering} shows the user clustering at time \emph{$t_{0}$}, there are four colors to represent users in each cluster and the number of users of each cluster is no more than 3. Additionally, the big circles are the position where the passive beamforming is aligned with the launch.
\setlength{\abovecaptionskip}{-0.2cm}
\par

\vspace{-0.2cm}
\subsection{Impact of IRSs}
we assume that the CSI is known and the attenuation process of the signal in the channel all follow the Rice distribution. Under this assumption, we analyze the sum rate over the total transmit power and the number of elements of the IRS. We randomly select one of clustering results, which is [1,3,1,2,3]. Besides, the resolution bits \emph{B} is set as 5.

\subsubsection{Sum Rate versus Total Transmit Power}
\begin{figure}[t!]
  \begin{center}
  \includegraphics[height=1.9in,width=2.7in]{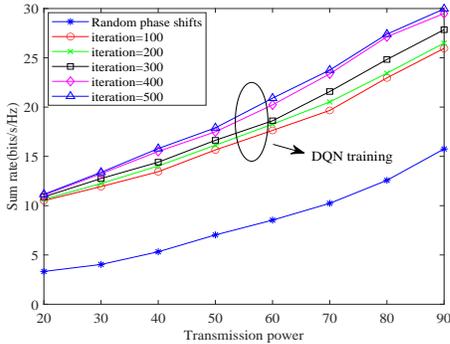}
  \caption{The sum rate versus total transmit power $\mathcal{P}$, \emph{K} = 25.}
  \label{sum rate versus power}
  \end{center}
  \end{figure}

Fig.~\ref{sum rate versus power} shows the relationship between transmit power and sum rate when the number of elements is fixed as 25. It is observed that the sum rate of all IRS-NOMA schemes increase with the increment of $\mathcal{P}$. It can also be observed that, the performance is capable of being improved over iterations. The proposed algorithm can converge after roughly a number of 500 iterations. Thus, the sum rate will not increase by adding more iterations after 500 iterations. Furthermore, we observe that the "Random phase shifts" approach performs worse when comparing to the proposed approach in terms of sum rate, which emphasizes the importance of phase shift design of the IRS.
\setlength{\abovecaptionskip}{-0.2cm}
\par

\subsubsection{Sum Rate versus the Number of IRS Elements}
\begin{figure}[t!]
  \begin{center}
  \includegraphics[height=1.9in,width=2.7in]{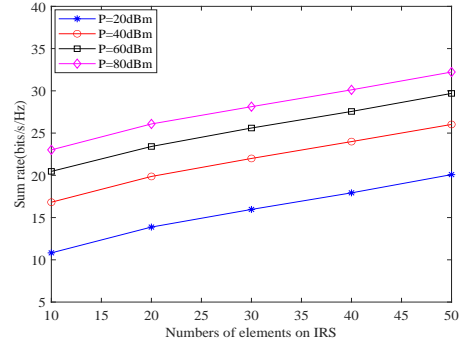}
  \caption{The sum rate versus IRS elements.}
  \label{sum rate versus IRS elements}
  \end{center}
  \end{figure}

In Fig.~\ref{sum rate versus IRS elements}, we compare the sum rate over the number of elements of the IRS under different transmission power from 20 dBm to 90 dBm. It shows that when the sum rate is improved with the increment of the number of elements. When the number of elements is with a large value, the growth rate of the sum rate over the number of elements slows down. Additionally, the gap grows wider when the transmission power is increased.
\setlength{\abovecaptionskip}{-0.2cm}
\par

\vspace{-0.2cm}
\subsection{Comparison between NOMA and OMA}
\begin{figure}[t!]
  \begin{center}
  \includegraphics[height=1.9in,width=2.7in]{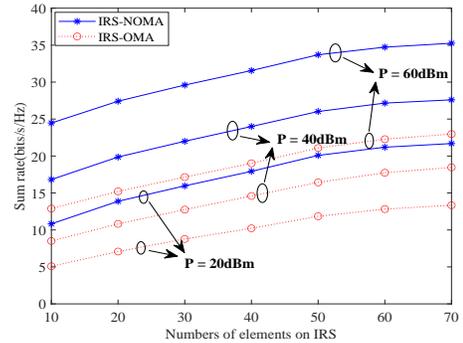}
  \caption{Comparison between NOMA and OMA.}
  \label{Comparison between NOMA and OMA}
  \end{center}
  \end{figure}

Finally, we compare the performance of the IRS-NOMA scheme to that of the IRS-OMA scheme. In the IRS-OMA scheme, the communication system follows the time division multiple access (TDMA) with the aid of the IRS. Under cases of three different transmit power 20 dBm, 40 dBm and 60 dBm, Fig.~\ref{Comparison between NOMA and OMA} characterizes the performance of both schemes. It can be observed that when all users are served by the BS simultaneously, when the transmit power is improved by 20dBm, the IRS-NOMA scheme outperforms the IRS-OMA scheme with more than 44.99$\%$ and 39.16$\%$.
\setlength{\abovecaptionskip}{-0.2cm}
\par

\vspace{-0.1cm}
\section{Conclusion}
\par
In this paper, we exploited the IRS-aided MISO-NOMA network. The sum rate maximization problem was formulated by jointly optimizing passive beamforming vector and power allocation efficient vector under QoS constraint. To tackle the problem formulated, three machine learning algorithms were proposed to predict user mobility, partition users into clusters and design the phase shift matrix, respectively. Numerical results were provided for demonstrating that the proposed IRS-NOMA scheme achieved significant performance gain compared to the IRS-OMA scheme. Additionally, properly choosing parameters in the proposed algorithms is capable of improve the training performance of neural networks.


%

\appendices

\vspace{-0.15cm}
\section*{Appendix~A: Proof of Proposition \ref{proposition 1}} \label{Appendix:A}
\vspace{-0.15cm}
\renewcommand{\theequation}{A.\arabic{equation}}
\setcounter{equation}{0}

\par
Suppose that there are two users \emph{$u_{m,p}$} and \emph{$u_{m,q}$} with optimal decoding order \emph{$R_{m,p \rightarrow m,q} \geq R_{m,q \rightarrow m,q}$}, it is simplified as
\vspace{-0.2cm}
\begin{align}
  \mid&\pmb{\Phi}_{m,p}\pmb{\omega}_{m}\alpha_{m,p}\mid^2(\sum\limits_{\lambda \neq p}\mid\pmb{\Phi}_{m,q}\pmb{\omega}_{m}\alpha_{m,\lambda}\mid^2+\mid\pmb{\Phi}_{m,q}\pmb{\Gamma}_{\gamma}\mid^2) \nonumber\\
  &\geq\mid\pmb{\Phi}_{m,q}\pmb{\omega}_{m}\alpha_{m,p}\mid^2(\sum\limits_{\lambda \neq p}\mid\pmb{\Phi}_{m,p}\pmb{\omega}_{m}\alpha_{m,\lambda}\mid^2+\mid\pmb{\Phi}_{m,p}\pmb{\Gamma}_{\gamma}\mid^2).
\end{align}
\par
And we add that the intra-cluster interference \emph{$\Omega = \mid\pmb{\Phi}_{m,q}\pmb{\omega}_{m}\alpha_{m,\lambda}\mid^2\Omega$} for user \emph{$u_{m,p}$} to both sides without inequality. Therefore, we can obtain that
\vspace{-0.2cm}
\begin{align}
R_{m,p \rightarrow m,q} \geq R_{m,q \rightarrow m,q} \geq R_{m,q \rightarrow m,\tilde{q}}.
\end{align}
Thus, we can get that the $R_{m,p \rightarrow m,q} \geq R_{m,q \rightarrow m,q}$ is the necessary condition of $R_{m,p \rightarrow m,q} \geq R_{m,q \rightarrow m,\tilde{q}}$.

\vspace{-0.15cm}
\section*{Appendix~B: Proof of Proposition \ref{proposition 2}} \label{Appendix:B}
\vspace{-0.15cm}
\renewcommand{\theequation}{B.\arabic{equation}}
\setcounter{equation}{0}

\par
An implicit variable \emph{$\varrho_{\lambda,m}$} is set to reflect the observation data \emph{$h_{p_{m}}$} from the m-th sub-model data. When the value of \emph{$\varrho_{\lambda,m} = 1$}, it means user \emph{$\lambda$} comes from \emph{m}th sub-model. Thus, the complete data is expressed as
\vspace{-0.2cm}
\begin{align}
C = (\pmb{S},\varrho_{\lambda,1},\varrho_{\lambda,2},\cdots,\varrho_{\lambda,M}).
\end{align}
\par
With observed data \emph{\pmb{S}} and unobserved data \emph{$\varrho_{\lambda,m}$}, we obtain the likelihood function of the complete data as follows
\vspace{-0.2cm}
\begin{align}
P(\pmb{S},\varrho|\pmb{\kappa}) = \prod\limits_{m=1}^{M} \Phi_{m}^{\lambda_{m}} \prod\limits_{\lambda=1}^{\Lambda} [\frac{1}{\sqrt{2\pi}\delta_{m}}exp(-\frac{(\tilde{h}_{\lambda} - C_{m})^2}{2\delta_{m}^{2}})]^{\varrho_{\lambda,m}},
\end{align}
where the \emph{$\lambda_{m}$} represents the number of data generated by the \emph{m}th sub-model out of \emph{$\Lambda$} observation data and satisfied the condition: \emph{$\lambda_{m} = \sum_{\lambda=1}^{\Lambda} \varrho_{\lambda,m}$} and \emph{$M = \sum_{m=1}^{M} \lambda_{m}$}, therefore, combined with the equation(29), the log-likelihood function of the final complete data can be expressed as
\vspace{-0.2cm}
\begin{align}
  &L(\pmb{\kappa}) = logP(\pmb{S}, \pmb{\varrho}|\pmb{\kappa}) \nonumber\\
  &= \sum\limits_{m=1}^{M}\{\lambda_{m}log\Phi_{m} + \sum\limits_{\lambda=1}^{\Lambda}\varrho_{\lambda,m}[log(\frac{1}{\sqrt{2\pi}\delta_{m}}) - \frac{(\tilde{h}_{\lambda} - C_{m})^2}{2\delta_{m}^{2}}]\}.
\end{align}
\par
The function refers to the expectation of the log-likelihood function \emph{$logP(\pmb{S}, \pmb{\varrho}|\pmb{\kappa})$} of the complete data given the observation data \emph{\pmb{S}} and the parameter \emph{$\pmb{\kappa}_{i}$} of the \emph{i}th iteration. The expected probability of calculation is the conditional probability distribution \emph{$logP(\pmb{S}, \pmb{\varrho}|\pmb{\kappa}_{i})$} of the hidden random variable \emph{$\pmb{\varrho}$}. So the Q function is given by
\vspace{-0.2cm}
\begin{align}
  Q(\pmb{\kappa},\pmb{\kappa}_{i}) &= \sum\limits_{m=1}^{M}\{\sum\limits_{\lambda=1}^{\Lambda}(E\varrho_{\lambda,m})log\Phi_{m} + \nonumber\\
  &\hspace*{3em}\sum\limits_{\lambda=1}^{\Lambda}(E\varrho_{\lambda,m})[log(\frac{1}{\sqrt{2\pi}\delta_{m}}) - \frac{(\tilde{h}_{\lambda}- C_{m})^2}{2\delta_{m}^{2}}]\},
\end{align}
where the conditional probability distribution of the implicit random variable \emph{$\kappa$} satisfied \emph{$logP(\pmb{\varrho},\pmb{S}|\pmb{\kappa}_{i}) = 1$}, thus the expectation of \emph{$E(\pmb{\varrho}_{\lambda,m}|\pmb{S},\pmb{\kappa}_{i})$} is calculated by
\vspace{-0.2cm}
\begin{align}
\tilde{\pmb{\varrho}}_{\lambda,m} = E(\pmb{\varrho}_{\lambda,m}|\pmb{S},\pmb{\kappa}_{i})=\frac{\Phi_{m}p(\pmb{S}_{\lambda}|\pmb{\kappa}_{m})}{\sum\limits_{m=1}^{M}\Phi_{m}p(\pmb{S}_{\lambda}|\pmb{\kappa}_{m})}.
\end{align}
\par
Derived from the above formula, we make the following equivalent transformations.
\vspace{-0.2cm}
\begin{align}
  Q(\pmb{\kappa},\pmb{\kappa}_{i}) = \sum\limits_{m=1}^{M}\{\lambda_{m}&log\Phi_{m} + \nonumber\\
  &\sum\limits_{\lambda=1}^{\Lambda}\tilde{\varrho}_{\lambda,m}[log(\frac{1}{\sqrt{2\pi}\delta_{m}}) - \frac{(\tilde{h}_{\lambda} - C_{m})^2}{2\delta_{m}^{2}}]\}.
\end{align}
\par
The next step is to obtain the parameter \emph{$\pmb{\kappa}_{i}$} of the \emph{i}th round, find the parameter \emph{$\pmb{\kappa}_{i+1}$} of the next iteration and make the function \emph{$Q(\pmb{\kappa},\pmb{\kappa}_{i})$} maximum, it is expressed as
\vspace{-0.2cm}
\begin{align}
\pmb{\kappa}_{i+1} = arg \max_{\pmb{\kappa}}Q(\pmb{\kappa},\pmb{\kappa}_{i}).
\end{align}
\par
It should be noted that we use \emph{$\chi_{\lambda,m}$} to denote the expectation of \emph{$\varrho_{\lambda,m}$}. Therefore, according to the EM derivation results, we can get the equation \eqref{23} to \eqref{24}.



\ifCLASSOPTIONcaptionsoff
  \newpage
\fi

\end{document}